# About transmission of thermal light through the surface of a body into the vacuum

M. V. Lebedev
Institute of Solid State Physics,
Russian Academy of Sciences,
Chernogolovka

The problem of the change of photon distribution of thermal light passing through the body surface into the vacuum can be formulated in a following way. Suppose we know the photon distribution $P_n^{in}$ of the thermal light inside some body occupying semi-infinite space. What will be the photon distribution $P_n^{out}$ of radiation transmitted through the surface of the body into the vacuum provided the energy transmission coefficient of the surface is $a$?

In a classic theory the light can be considered as electromagnetic wave and the energy transmitted into the vacuum will be:

$$E^{out} = aE^{in} \qquad (1)$$

In quantum mechanical theory the energy of electromagnetic field in one of its energetic states consists of an integer number of quanta, that's why different possibilities arise in this case. It's clear that for mean energies the expression (1) still holds:

$$\langle E^{out} \rangle = a \langle E^{in} \rangle \qquad (2)$$

But this equation can be fulfilled in various ways. The most natural one looks as the assumption that each photon passes through the boundary totally independent of others with a probability $a$. It is straightforward to write the expression for the photon distribution for the transmitted light in this case [1-3]:

$$P_n^{out} = \frac{(a\langle n_{in} \rangle)^n}{(1 + a\langle n_{in} \rangle)^{n+1}} \qquad (3)$$

Another possible assumption is that the probability $a$ describes the transmission of a light state $P_n^{in}$ as a whole. In other words, in the case electromagnetic field in a state $P_n^{in}$ falls onto the surface, all $n$ photons are transmitted with probability $a$, and reflected with the probability $1-a$, but the cases when some of them are transmitted and some reflected are never observed. This assumption leads to:

$$P_n^{out} = aP_n^{in}, \quad n \geq 1$$
$$(1 - P_0^{out}) = a(1 - P_0^{in}) \qquad (4)$$

Physically this means that photons are transmitted through the boundary not independently but conserving, to a great extent, the correlations coming from the initial state of a light field. One can say that Eq. (3) corresponds to the transmission of independent particles and Eq. (4) to the transmission of quantized waves. It is clear that the statement that photons pass through the boundary independently is an additional assumption which does not follows from the Kirchhoff's law immediately because the assumption (4) fulfils this law too. Moreover, one can consider some other distributions of transmitted photons which fulfill the Kirchhoff's law. So we see that the solution of the problem of the photon distribution for a thermal light emitted with a heated body is not obvious and one has to find some additional criteria which should be fulfilled apart from the Kirchhoff's law. It will be shown in what follows that the assumption (3) contradicts the expression for the fluctuation of thermal light found by A. Einstein in 1909 [4].

## The Kirchhoff's law and photon distribution of thermal light

According to the Kirchhoff's law the energy spectral density of radiation emitted by a heated body can be expressed as:
$$\rho(\omega,T) = a(\omega) f(\omega,T) \tag{5}$$
where $f(\omega,T)$ is a universal function of frequency and temperature and $a(\omega)$ depends on the body properties. The function $f(\omega,T)$ is the energy spectral density of a black body for which by definition $a(\omega) \equiv 1$ holds. A black body is usually modelled experimentally as a cavity kept at some temperature with a small hole in a wall of the cavity from which thermal radiation is emitted. Let us show that the spectral density of thermal radiation emitted from such a hole will be given with Eq. (5) with $a(\omega) \equiv 1$.

Consider some solid body at temperature $T$ having a flat surface. Suppose a flat ideal mirror is brought to the surface of the body on some small distance but still large enough as compared to the wavelength of light. The surfaces of the mirror and the body form a cavity which will be filled with thermal radiation. For a large enough surface the energy losses through the edges can be neglected and for the energy spectral density of radiation emitted from a small hole in the mirror one can write:
$$\rho(\omega,T) = af + (1-a)af + (1-a)^2 af + \ldots = af\{1 + (1-a) + (1-a)^2 + \ldots\} = f \tag{6}$$
This radiation will thus have a spectral density of emitted energy just as the ideal black body. The Kirchhoff's law states explicitly that thermal radiation is attenuated by a transmission coefficient of the boundary when going out from the body into the vacuum. Function $f(\omega,T)$ is the well known Plank's function:
$$f(\omega,T)d\omega = \frac{\hbar\omega^3}{\pi^2 c^3} \frac{1}{e^{\frac{\hbar\omega}{kT}} - 1} d\omega = \langle n \rangle \hbar\omega \frac{\omega^2}{\pi^2 c^3} d\omega \tag{7}$$

where $\langle n \rangle = \dfrac{1}{e^{\frac{\hbar\omega}{kT}} - 1}$ is the mean number of quanta occupying a single phase space sell of the ideal photon gas, $\hbar\omega$ is the energy of a quantum and $\dfrac{\omega^2}{\pi^2 c^3} d\omega$ - the number of phase space sells in a frequency interval $d\omega$. The Kirchhoff's law states thus that the mean number of quanta per unit sell is reduced for thermal radiation in accordance with:
$$\langle n_{out} \rangle = a \langle n_{in} \rangle \tag{8}$$

where $\langle n_{out} \rangle$ and $\langle n_{in} \rangle$ are the mean numbers of quanta per phase space sell for radiation outside and inside of the radiating body. The probability $P_n^{in}$ to find just $n$ photons in a phase space sell is for thermal radiation inside the body given with a well known Bose-Einstein distribution:

$$P_n^{in} = \frac{\langle n_{in} \rangle^n}{(1+\langle n_{in} \rangle)^{n+1}} \qquad (9)$$

One can choose between Eqs. (3) and (4) for $P_n^{out}$ calculating the fluctuation of thermal light energy inside the cavity and compare the results with the result found in 1909 by Einstein [4] on the basis of general thermodynamic considerations:

$$\langle n_{in}^2 \rangle = \langle n_{in} \rangle + 2\langle n_{in} \rangle^2 \qquad (10)$$

The propagation of light through the body boundary into the vacuum is not a measurement in a quantum mechanical sense and should therefore be treated in terms of a free evolution of light states in a space with given boundary conditions. Let us denote the probability to find just $n$ photons in a phase space sell for a transmitted through the boundary light as $Q_n$ to distinguish it from $P_n$ - the Bose-Einstein probability. Considering successive reflections of a given light state from the mirror and the body surface we get probabilities $Q_n^i$ to find this state after $i$ reflections. For the total probability, which should be evidently the Bose-Einstein probability, holds:

$$P_n = \sum_{i=0}^{\infty} Q_n^i \qquad (11)$$

We can now verify the hypotheses (3) and (4) by calculating $\langle n^2 \rangle$ and comparing the result obtained with Eq.(10).

## Calculation of $\langle n^2 \rangle$ on the basis of the assumption of independent photon transmission trough the boundary between the body and the vacuum

Starting from Eq. (3) we get:

$$\langle n^2 \rangle = \sum_{n=0}^{\infty} n^2 \sum_{i=0}^{\infty} Q_n^i = \sum_{n=0}^{\infty} n^2 \sum_{i=0}^{\infty} \frac{[a(1-a)^i \langle n \rangle]^n}{[1+a(1-a)^i \langle n \rangle]^{n+1}} = \langle n \rangle + 2\sum_{i=0}^{\infty} \{a(1-a)^i \langle n \rangle\}^2 \qquad (12)$$

Summing up the geometric sequence in Eq. (12) gives:

$$\langle n^2 \rangle = \langle n \rangle + \frac{2a\langle n \rangle^2}{2-a} \qquad (13)$$

Eq. (13) obviously tends to Eq. (10) when $a \to 1$.

## The hypothesis of wave-like transmission of photons through the boundary

The assumption (4) gives:

$$\langle n^2 \rangle = \sum_{n=0}^{\infty} n^2 \sum_{i=0}^{\infty} a(1-a)^i P_n = \langle n \rangle + 2\langle n \rangle^2 \quad (14)$$

This result does not depend on the surface transparency. Comparison of (13) and (14) shows that one should prefer the hypothesis of wave-like transmission (4). This fact cannot of course be considered as a strong evidence, because one cannot exclude a possibility of existing another hypothesis consistent both with the Kirchhoff's law and with expression (10), but the close connection of (4) with the wave picture of light makes the validity of this hypothesis very likely.

## References

1. L.Mandel and E.Wolf, Optical coherence and quantum optics, 1995, Cambridge University Press.
2. D.N. Klyshko, Sov.Phys.JETP, 1986, 63 (4), 682-687
3. D.N. Klyshko, Phys. Usp.,1996, **39,** 573–596
4. A. Einstein, Phys. Zs., 1909, **10**, 185-193